\documentclass[aps,prd,10pt,twocolumn,superscriptaddress,nofootinbib,noshowpacs,preprintnumbers,hyperref,floatfix,reprint]{revtex4-1}

\pdfoutput=1

\usepackage{amsmath}
\usepackage{amsfonts}
\usepackage{amssymb}
\usepackage{bbold}
\usepackage{epsfig}
\usepackage{graphicx}
\usepackage{bm}
\usepackage{array}
\usepackage{hyperref}
\usepackage{listings}
\usepackage{color}
\usepackage[normalem]{ulem}
\usepackage{cancel}
\usepackage{multirow}

\newcommand{\eref}[1]{Eq.~(\ref{#1})}
\newcommand{\fref}[1]{Fig.~\ref{#1}}

\newcommand{\tref}[1]{Table~\ref{#1}}

\newcommand{\Ar}{^{40}\hspace{-0.7mm}{\rm Ar}}
\newcommand{\K}{^{40}\hspace{-0.7mm}{\rm K}}

\usepackage{tikz,xcolor,hyperref}

\definecolor{lime}{HTML}{A6CE39}
\DeclareRobustCommand{\orcidicon}{\hspace{-1mm}
	\begin{tikzpicture}
	\draw[lime, fill=lime] (0,0) 
	circle [radius=0.16] 
	node[white] {{\fontfamily{qag}\selectfont \tiny \,ID}};
	\draw[white, fill=white] (-0.0525,0.095) 
	circle [radius=0.007];
	\end{tikzpicture}
	\hspace{-3mm}
}

\foreach \x in {A, ..., Z}{\expandafter\xdef\csname orcid\x\endcsname{\noexpand\href{https://orcid.org/\csname orcidauthor\x\endcsname}
			{\noexpand\orcidicon}}
}

\begin{document}

\title{Triangulating Black Hole Forming Stellar Collapses through Neutrinos}% 

\author{Lila Sarfati\orcidA{}}%
%\email{lila.sarfati@ens.psl.eu}
\affiliation{%
 Niels Bohr International Academy and DARK, Niels Bohr Institute, University of Copenhagen, Blegdamsvej 17, 2100, Copenhagen, Denmark
}
\affiliation{D{\'e}partement de Physique de l'\'Ecole Normale Sup{\'e}rieure - PSL,
45 rue d'Ulm, 75230, Paris Cedex 05, France}
\author{Rasmus S.~L.~Hansen\orcidB{}}%
%\email{rslhansen@nbi.ku.dk}
\affiliation{%
 Niels Bohr International Academy and DARK, Niels Bohr Institute, University of Copenhagen, Blegdamsvej 17, 2100, Copenhagen, Denmark
}

\author{Irene Tamborra\orcidC{}}%
%\email{tamborra@nbi.ku.dk}
\affiliation{%
 Niels Bohr International Academy and DARK, Niels Bohr Institute, University of Copenhagen, Blegdamsvej 17, 2100, Copenhagen, Denmark
}%

\date{\today}

\begin{abstract}
In the event of a  black hole (BH) forming stellar collapse, 
the neutrino signal should terminate abruptly at the moment of BH formation, after a phase of steady accretion.  Since neutrinos are expected to reach Earth hours before the electromagnetic signal, the combined detection of the neutrino burst through multiple neutrino telescopes could allow to promptly determine the angular location of a nearby  stellar collapse in the sky with high precision. 
In this paper, we contrast the  triangulation   pointing procedure  that relies on the rise time of the neutrino curve, often considered in the literature, to the one that takes advantage of the   cutoff of the neutrino curve at the moment of BH formation. By forecasting the neutrino signal expected in the IceCube Neutrino Observatory, Hyper-Kamiokande and DUNE, we devise  a strategy to optimize the  identification of the rise and  cutoff time of the neutrino curve. We show that the triangulation method developed by employing  the end tail of the neutrino curve  allows to achieve at least  one order of magnitude improvement in the pointing precision for a galactic burst, while being insensitive to the neutrino mixing scenario. The triangulation   pointing   method based on the cutoff of the neutrino curve will also guarantee a better performance for BH forming collapses occurring beyond our own Galaxy. 
\end{abstract}

\maketitle

%%%%%%%%%%%%%%%%%%%%%%%%%%%%%%%%%%%%%%%%%%%%%%%%%%
%%%%%%%%%%%%%%%%%%%%%%%%%%%%%%%%%%%%%%%%%%%%%%%%%%
\section{Introduction}
%%%%%%%%%%%%%%%%%%%%%%%%%%%%%%%%%%%%%%%%%%%%%%%%%%
%%%%%%%%%%%%%%%%%%%%%%%%%%%%%%%%%%%%%%%%%%%%%%%%%%

Being copiously produced, neutrinos carry key information about and strongly affect the physics of the latest stages of  life of  stars with a zero-age main sequence mass  between $10$ and $150\ M_\odot$~\cite{Horiuchi:2018ofe,2019ARNPS..69..253M,Mirizzi:2015eza,1971ApJ...163..209W}.  For core-collapse supernova explosions, the bounce of the electron-degenerate iron core induces a shock wave in the infalling stellar matter that is  revived by neutrinos, after stalling~\cite{1990RvMP...62..801B,1999ApJ...522..413F,1966ApJ...143..626C,1971ApJ...163..221C,2002RvMP...74.1015W}. 
However, if the explosion mechanism fails and matter continues to accrete onto the transiently stable proto-neutron star,  a stellar-mass black hole (BH) forms.

Recent theoretical work~\cite{Ertl:2015rga,Sukhbold:2015wba,OConnor:2010moj,Sotani:2021kvj,2020ApJ...890..127C} as well as the observation of disappearing red giants~\cite{Adams:2016ffj,Adams:2016hit,Neustadt:2021jjt,Smartt:2015sfa,Gerke:2014ooa,Villarroel:2019bky} hint towards the possibility that BH formation may occur for $10$--$30\%$ of all massive stars undergoing core collapse, and possibly more. Such collapses are  expected to emit a faint electromagnetic signal due to the stripping of the hydrogen envelope~\cite{Kochanek:2013yca,Lovegrove:2013ssa,1980Ap&SS..69..115N}, hence neutrinos (and probably gravitational waves) could be  the only messengers of such transient events.  

It is crucial to  hunt for BH forming collapses through neutrinos~\cite{Scholberg:2017czd,Mirizzi:2015eza}. 
The smoking gun of BH formation~\cite{1996ApJ...458..680B,Keil:1995hw,1995ApJ...448..797G,Baumgarte:1996iu,Zha:2021fbi,daSilvaSchneider:2020ddu,OConnor:2010moj,Gullin:2021hfv} would be the relatively abrupt termination of the observable neutrino curve~\footnote{Neutrino curve is  used  in analogy to light curve in this work.} after a period of post-bounce accretion lasting for a few hundreds  milliseconds up to several seconds~\cite{1988ApJ...334..891B,OConnor:2010moj,Kresse:2020nto}. 

The early detection of neutrinos is especially important in aiding the prompt reconstruction of   directional information about the stellar collapse.
To this purpose, the neutrino signal could be exploited for pointing and triangulation~\cite{Beacom:1998fj, Tomas:2003xn,Scholberg:2009jr,Muhlbeier:2013gwa,Fischer:2015oma,Brdar:2018zds,Hansen:2019giq,Linzer:2019swe,SNEWS:2020tbu,Nakamura:2016kkl,Adams:2013ana}. In particular, in water Cherenkov neutrino detectors, such as Super-Kamiokande~\cite{Super-Kamiokande:2007zsl} and the upcoming Hyper-Kamiokande~\cite{Hyper-Kamiokande:2018ofw}, the electron-neutrino elastic scattering can be employed to determine the angular location of the stellar collapse, allowing  to achieve a few-degree pointing precision~\cite{Super-Kamiokande:2016kji,Beacom:1998fj}. The  upgrade of Super-Kamiokande with gadolinium foresees an increased tagging efficiency that will  improve the pointing precision by $50\%$, while a sub-degree precision could be potentially reached with Hyper-Kamiokande~\cite{Tomas:2003xn}. A similar performance is expected from DUNE~\cite{Nakamura:2016kkl,Bueno:2003ei}.
 Triangulation by computing the time delays of the neutrino signal from standard core collapse supernovae among
various neutrino telescopes  does not allow to obtain such a precision. However, for BH forming collapses, an improved pointing precision of triangulation is guaranteed  because of the higher neutrino event statistics achievable  with respect to standard core-collapse supernovae~\cite{Muhlbeier:2013gwa,Brdar:2018zds,Hansen:2019giq}.

In this paper, we investigate the possibility of improving on the determination of the angular location of a BH forming collapse. We work  under the assumption of a negligible or absent supersonic accretion flow in the surroundings of the collapsing proto-neutron star, leading to the formation of a naked BH.  In fact, in such a scenario, one should expect a sharp drop of the neutrino signal~\cite{1996ApJ...458..680B,Keil:1995hw,1995ApJ...448..797G,Baumgarte:1996iu}, with no smearing of the tail determined by the coherent scattering of neutrinos off nuclei, possibly occurring in the event of a non-negligible accretion flow~\cite{Gullin:2021hfv}.  We show that, by focusing on the tail of the neutrino curve instead of on its bounce time to compute the time delays, one can achieve at least an order of magnitude better determination of the angular location of the stellar collapse because of the  larger event rate expected at the end of the accretion phase, which is more easily distinguishable from the background signal. 

This paper is organized as follows. In Sec.~\ref{sec:input}, we introduce the  neutrino emission properties  for our BH forming collapse model. The main features of the  neutrino telescopes adopted in this work are outlined in Sec.~\ref{sec:detectors}. The method adopted to determine the timing of the neutrino signal is introduced in Sec.~\ref{sec:timing}, a comparison between the determination of the time delay  among pairs of detectors is discussed by employing the  rise time as well as the cutoff time of the neutrino curve. In Sec.~\ref{sec:triangulation}, our findings on the triangulation of the neutrino signal are presented. The dependence our findings  on the distance at which the BH forming collapse occurs can be found in Sec.~\ref{sec:distance}, followed by a discussion on our main results and caveats in Sec.~\ref{sec:discussion} and concluding remarks in Sec.~\ref{sec:conclusions}.

%%%%%%%%%%%%%%%%%%%%%%%%%%%%%%%%%%%%%%%%%%%%%%%%%%
%%%%%%%%%%%%%%%%%%%%%%%%%%%%%%%%%%%%%%%%%%%%%%%%%%
\section{Neutrino emission properties for black hole forming collapses}
\label{sec:input}
%%%%%%%%%%%%%%%%%%%%%%%%%%%%%%%%%%%%%%%%%%%%%%%%%%
%%%%%%%%%%%%%%%%%%%%%%%%%%%%%%%%%%%%%%%%%%%%%%%%%%

The neutrino emission properties for our benchmark BH forming collapse  are extracted from a one-dimensional (1D) spherically symmetric hydrodynamical simulation of BH forming collapse from the Garching group~\cite{Garc:SN,Mirizzi:2015eza}:  the $40\ M_\odot$  model with slow accretion rate  (s40c) for which the gravitational instability occurs after $2100$~ms.  The employed high density nuclear equation of state  is the Lattimer and Swesty one with nuclear incompressibility modulus $K=220$~MeV~\cite{Lattimer:1991nc}. We refer the interested reader to Sec.~2.5 of Ref.~\cite{Mirizzi:2015eza} for more details on this model.

 It is worth to be noted that BH forming collapses have been simulated in multi-D, see e.g.~Refs.~\cite{Chan:2017tdg,Pan:2017tpk,Kuroda:2018gqq,Walk:2019miz,Pan:2020idl}, and 3D  hydrodynamical simulations with sophisticate neutrino transport exhibit strong signatures of hydrodynamical instabilities in the neutrino curve~\cite{Walk:2019miz}. However, since we are interested in the rise time of the neutrino curve as well as in its drop, and these features are not affected by the dimensionality of the simulation, a 1D model suits our purpose well.

Dedicated simulations following the dynamical implosion of the transiently stable neutron star into a naked BH  foresee two different scenarios: either   the naked or low accreting neutron star  collapses into a BH by cooling or a softening of the nuclear equation of state leads to a phase transition where exotic states of matter form~\cite{1996ApJ...458..680B,Keil:1995hw,1995ApJ...448..797G,Baumgarte:1996iu,Zha:2021fbi,daSilvaSchneider:2020ddu}. In the first case, the  neutrino signal terminates abruptly at the moment of BH formation, whereas in the second scenario  a gradual decrease of the neutrino signal is foreseen as more and more matter approaches the event horizon. 

An alternative scenario with respect to the one of naked BH formation described above foresees a substantial  accretion flow surrounding the collapsing protoneutron star. Under the assumption of spherical accretion,   the arrival time of neutrinos may be  affected by the coherent scattering of neutrinos in the infalling medium surrounding the BH~\cite{Gullin:2021hfv};  this would  result in a longer tail of the neutrino curve (so-called echo).

Triangulation through neutrinos could be carried out both in the case of the formation of a naked BH and when a non-negligible accretion affects the tail of the neutrino curve. However, in this work, we choose to focus on the scenario of naked BH formation since we are interested in quantitatively exploring the best circumstances under which triangulation could be carried out for BH forming collapses.

Our benchmark  model  has an instantaneous cutoff of the neutrino emission properties at the time of BH formation. Since we are interested in the tail of the neutrino curve, we assume that the signal predicted in Ref.~\cite{Baumgarte:1996iu} is  representative of naked BH formation scenarios. In fact, Ref.~\cite{Baumgarte:1996iu} followed the hyperonization of the cooling core during the collapse of a protoneutron star to a BH. 

We  take into account  a linear tail in the neutrino signal due to the increasing gravitational redshift lasting for approximately $0.5{\;\rm ms}$~\cite{Baumgarte:1996iu} 
The non-radial neutrino emission may contribute to soften  the cutoff of the neutrino curve~\cite{Wang:2021elf}, adding an  exponential contribution at the end of the neutrino tail; we  assume that  this effect  is also enclosed in  the representative  tail of the neutrino curve. In addition, the contribution from neutrinos on unstable circular orbits is expected to decay exponentially with a decay time of $\mathcal{O}(0.05){\;\rm ms}$~\cite{Beacom:2000qy,Baumgarte:1996iu}, however  this sub-leading component is neglected. We will return on the impact of our modeling assumptions in Sec.~\ref{sec:discussion}.

\begin{figure*}
  \centering
  \includegraphics[width=0.9\textwidth]{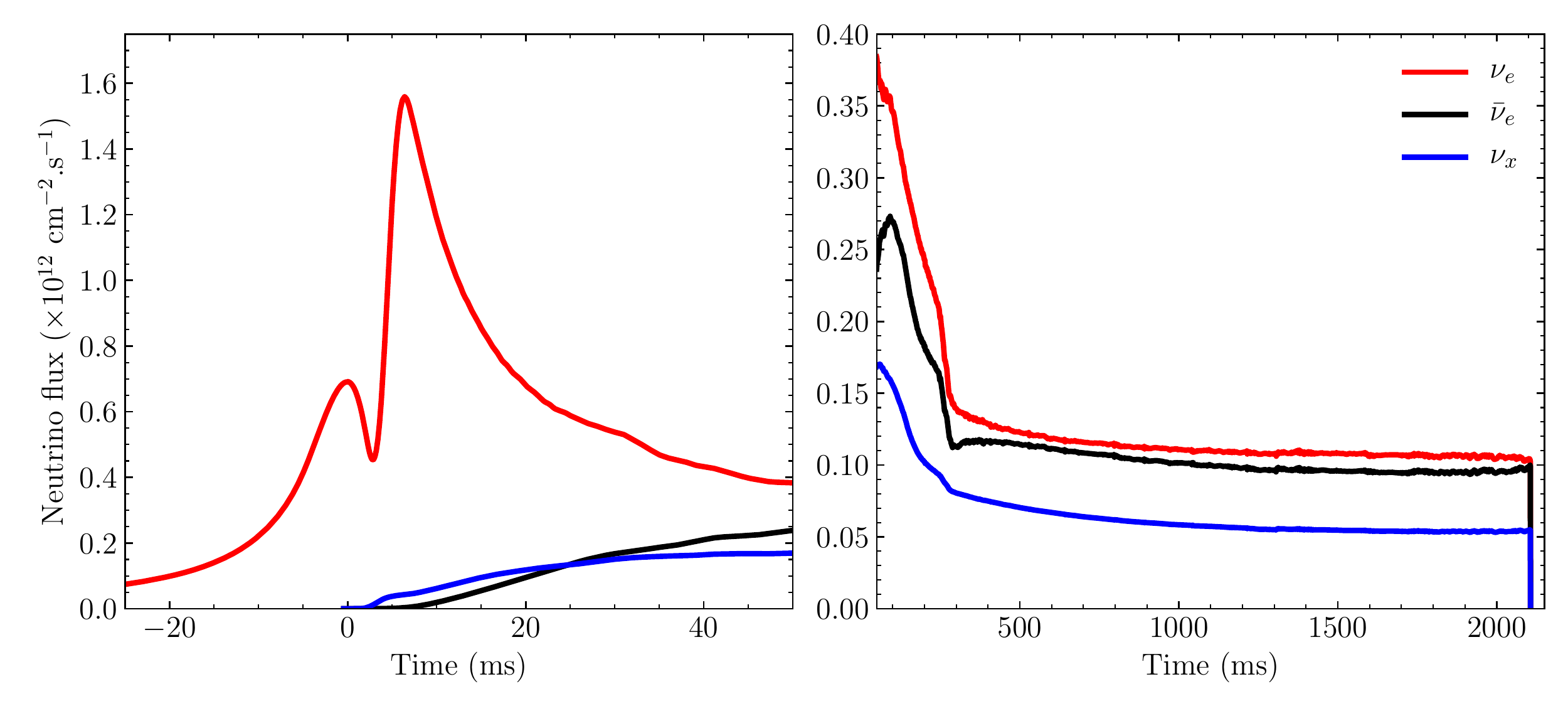}
  \caption{Neutrino flux in the absence of neutrino mixing as a function of  time for the 1D spherically symmetric BH forming collapse model  adopted in this work (model s40c of Ref.~\cite{Mirizzi:2015eza} with  tail modelled following  Ref.~\cite{Baumgarte:1996iu}). The $\nu_e$ flux is plotted in red, $\bar\nu_e$ in black and $\nu_x = \nu_\mu=\nu_\tau = \bar\nu_x$ in blue. The left panel shows the the  pre-bounce neutrinos and the  deleptonization burst. The right panel displays the  long-lasting accretion phase followed by a tail at $\sim 2100\;{\rm ms}$.}
  \label{fig:SNmodels}
\end{figure*}
By relying on the inputs from our benchmark  BH stellar collapse model, for each flavor $\nu_\beta$ (with $\beta= e, \mu$ or $\tau$ and $\nu_{\mu}=\nu_\tau = \nu_x = \bar\nu_x$ hereafter) and post-bounce time $t$, we compute the neutrino differential flux for a stellar collapse occurring at a distance $d$ as follows:
\begin{equation}
\label{eq:flux}
F_{\nu_\beta}(E,t) = \frac{L_{\nu_\beta}(t)}{4 \pi d^2} \frac{\phi_{\nu_\beta}(E,t)}{\langle E_{\nu_\beta}(t)\rangle}\ ,
\end{equation}
with $E$ being the neutrino energy, $L_{\nu_\beta}$  the  luminosity of the neutrino flavor $\nu_\beta$, and $\langle E_{\nu_\beta}\rangle$  the neutrino mean energy. The neutrino energy distribution is~\cite{Keil:2002in,Tamborra:2012ac}: 
\begin{equation}
\phi_{\nu_\beta}(E,t)= \xi_{\nu_\beta}(t) \left(\frac{E}{\langle E_{\nu_\beta}(t)\rangle}\right)^{\alpha_{\nu_\beta}(t)} \exp \left[- \frac{(1+\alpha_{\nu_\beta}) E}{\langle E_{\nu_\beta}(t)\rangle} \right]\ ;
\end{equation}
where $\xi_{\nu_\beta}(t)$ is  a normalization factor [$\int dE \phi_{\nu_\beta}(E,t) = 1$] and the pinching factor is defined such that
\begin{equation}
\frac{\langle E_{\nu_\beta}(t)^2 \rangle}{\langle E_{\nu_\beta}(t)\rangle^2} = \frac{2+\alpha_{\nu_\beta}(t)}{1+\alpha_{\nu_\beta}(t)}\ ,
\end{equation}
with $\langle E_{\nu_\beta}^2 \rangle$ being the second energy moment. 

 Figure~\ref{fig:SNmodels} shows the temporal evolution of the  flux (Eq.~\ref{eq:flux}) of neutrinos and antineutrinos of all flavors for our BH forming collapse occurring at $10$~kpc from the observer. After the neutronization burst, we can see the long-lasting accretion phase characteristic of our BH forming collapse model together with the abrupt termination of the neutrino signal at the moment of BH formation.

While propagating in the stellar envelope, neutrinos undergo flavor mixing~\cite{Mirizzi:2015eza}. In addition to the well known Mikheyev-Smirnov-Wolfenstein  (MSW) effect~\cite{Mikheev:1986if,1985YaFiz..42.1441M,1978PhRvD..17.2369W}, relevant at large distances from the core, neutrino-neutrino interaction is important in the innermost region, and especially in the  neutrino decoupling region~\cite{Tamborra:2020cul,Duan:2010bg,Chakraborty:2016yeg,Mirizzi:2015eza}.  In addition, if the matter background should have significant fluctuations,  flavor conversion physics would be further affected, as it happens in the presence of turbulence~\cite{Kneller:2017lqg,Patton:2014lza}.  

The modeling of the neutrino propagation and its flavor evolution in the dense stellar core is an area of active research due to the conceptual and numerical challenges that it entails, see e.g.~Refs.~\cite{Mirizzi:2015eza,Duan:2010bg,Tamborra:2020cul} for dedicated discussions.   In order to factor in any uncertainty in the neutrino signal linked to flavor mixing, in this paper, we rely on  two extreme mixing scenarios to   illustrate the maximum variability of the neutrino signal. The first scenario that we consider is the absence of
 neutrino mixing (referred to as ``no-mix''), such that  the neutrino signal is not affected by flavor mixing.
The second extreme scenario is the one of full flavor mixing (referred to as ``full-mix''), 
such that the flux of $\nu_e$ and the one of $\nu_x$ are swapped under the assumption of full flavor conversion; and a similar assumption holds for  $\bar{\nu}_e$ and $\bar{\nu}_x$.

%%%%%%%%%%%%%%%%%%%%%%%%%%%%%%%%%%%%%%%%%%%%%%%%%%
%%%%%%%%%%%%%%%%%%%%%%%%%%%%%%%%%%%%%%%%%%%%%%%%%%
\section{Neutrino telescopes}
\label{sec:detectors}
%%%%%%%%%%%%%%%%%%%%%%%%%%%%%%%%%%%%%%%%%%%%%%%%%%
%%%%%%%%%%%%%%%%%%%%%%%%%%%%%%%%%%%%%%%%%%%%%%%%%%

In order to explore the  pointing precision for a BH forming collapse by relying on the final tail of the neutrino signal or on the bounce time, we consider  three of the largest existing and upcoming neutrino telescopes: Hyper-Kamiokande (HK), the Deep Underground Neutrino Experiment (DUNE), and the IceCube Neutrino Observatory (IC). 
In addition to their size, the geographic distribution of these detectors provides a powerful setup for triangulating BH forming collapses through neutrinos. The main features of these detectors are introduced in this section.

\subsection{Hyper-Kamiokande}

Hyper-Kamiokande~\cite{Hyper-Kamiokande:2018ofw} is an upcoming water Cherenkov neutrino telescope, which will be located in Japan and will be the successor of Super-Kamiokande, well known for its sensitivity to  supernova neutrinos due to its high photocoverage~\cite{Super-Kamiokande:2007zsl}. 
Following Ref.~\cite{Linzer:2019swe}, we assume that the total mass of Hyper-Kamiokande will be $374$~kt with detection efficiency  $\epsilon_{\mathrm{HK}} = 100\%$~\cite{Super-Kamiokande:2002weg}. 
The angular coordinates of the geographical location of Hyper-Kamiokande are reported in Table~\ref{tab:coordinates}.
\begingroup
\setlength{\tabcolsep}{5pt}
\renewcommand{\arraystretch}{1.2}
\begin{table}
\caption{Angular coordinates of the three neutrino detectors employed in this work~\cite{Linzer:2019swe, IceCube:2016zyt}.  }
\begin{tabular}{l c c}
\hline\hline
Detector & Latitude (deg) & Longitude (deg)\\
\hline
HK   & $36.4$  &$137.3$\\
DUNE  & $44.4$ & $-103.8 $\\
IC   & $-90.0$ & $0.0$
\\
\hline\hline
\end{tabular}
\label{tab:coordinates}
\end{table}
\endgroup

\begin{figure}[ht]
  \centering
    \includegraphics[width=0.45\textwidth]{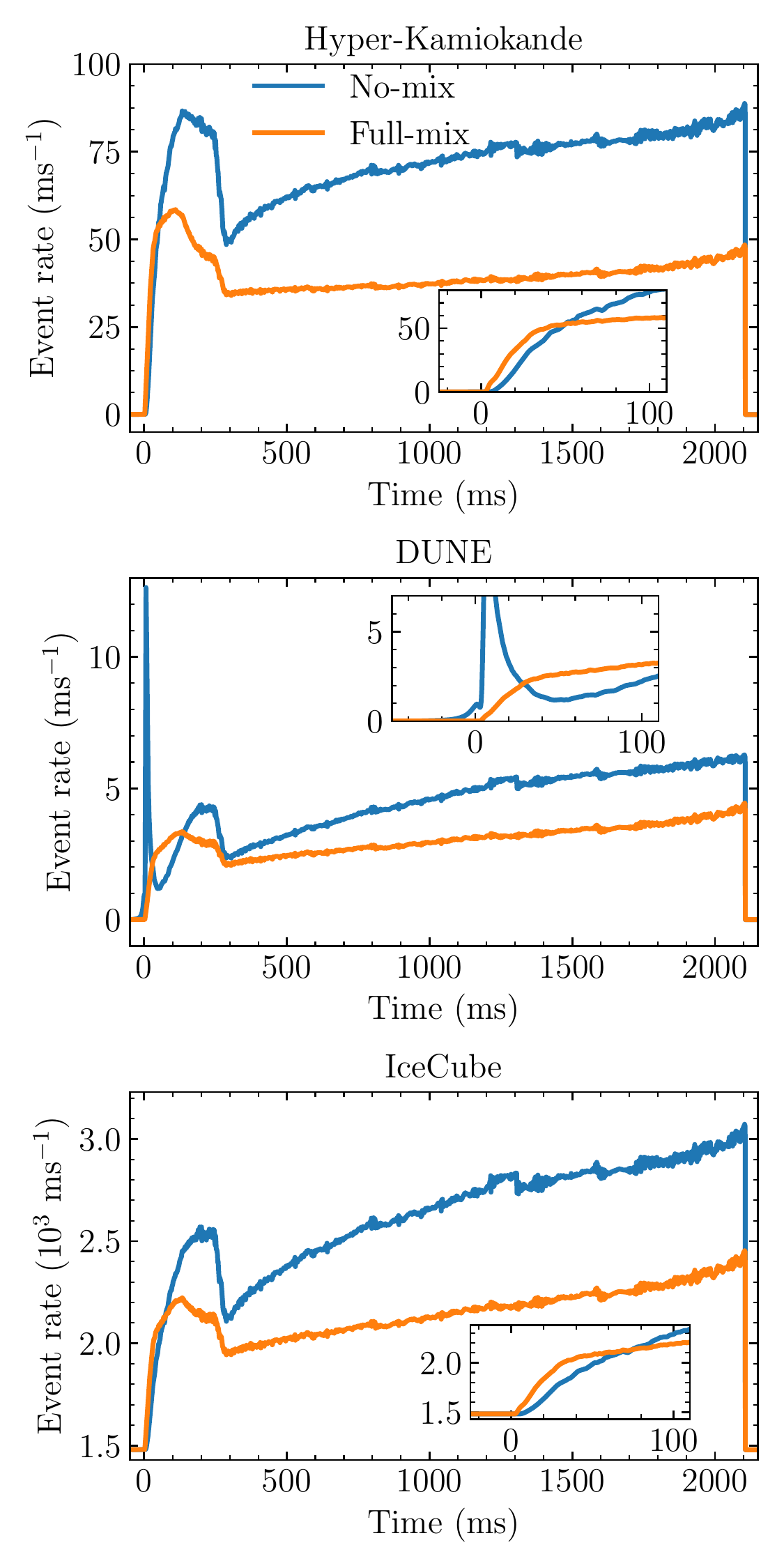}
  \caption{Expected neutrino event rate  as a function of the post-bounce time for a BH forming collapse located a $10$~kpc. The event rate is plotted for Hyper-Kamiokande,  DUNE and IceCube from top to bottom, respectively and for the two extreme mixing scenarios considered in this work.  The detector background has been plotted for each detector, but it is only visible for IC in the form of an offset of the event rate. In each panel, the inset plots highlight the rise of the neutrino curve.}  
  \label{fig:rate}
\end{figure}

The main channel of neutrino detection is inverse beta decay (IBD): $\bar{\nu}_e + p \rightarrow e^+ + n$. Other sub-dominant channels are neutrino-electron elastic scattering and neutrino-oxygen charged current interactions, but these are not considered in this work. The neutrino event rate is computed by folding the neutrino differential flux defined in Eq.~\ref{eq:flux} with the IBD differential cross section $\sigma_{\mathrm{IBD}}$~\cite{Strumia:2003zx}.
The kinematic threshold for IBD is $E_{\rm thr} = m_{e^+} + \Delta m_{np} = 1.804 {\; \rm MeV}$, while Hyper-Kamiokande is designed to have a threshold of $\sim 3{\;\rm MeV}$ for the positron kinetic energy. However, because of the typical neutrino energy from BH forming collapses and 
the fact that the background is large  at small energies,  we use  $E_{e, \rm thr} = 7{\;\rm MeV}$~\cite{Super-Kamiokande:2016kji} as threshold for the positron kinetic energy. This corresponds to a threshold on the neutrino energy of $8.293{\;\rm MeV}$. The resultant event rate for our BH forming collapse is 
\begin{equation}
\label{eq:HKrate}
R_{\mathrm{HK}} = N_p \epsilon_{\mathrm{HK}} \int_{E_{e, \rm thr}} dE_e \int dE F_{\bar\nu_e}(E) \sigma_{\mathrm{IBD}}(E_e, E)\ , 
\end{equation}
where $N_p$ is the number of protons in Hyper-Kamiokande and $E_e$ is the positron energy. The neutrino event rate expected in Hyper-Kamiokande for our BH forming collapse benchmark model is shown in the top panel of Fig.~\ref{fig:rate} as a function of time.
The ``no-mix'' case shows higher event numbers than the ``full-mix'' one due to the higher flux of  $\bar{\nu}_e$ compared to $\nu_x$ as seen in \fref{fig:SNmodels}. This is a feature of the accretion phase, although the exact relative evolution as a function of time for the two mixing scenarios depends on the mass of the supernova model, see e.g.~Refs.~\cite{Nagakura:2020qhb,Seadrow:2018ftp,Serpico:2011ir,Tamborra:2014hga,Walk:2019miz}.

Concerning the background, we  base our estimate on Ref.~\cite{Linzer:2019swe} which quote $0.1$~s$^{-1}$ events per $20$~kt. For Hyper-Kamiokande,  this corresponds to a total background rate of $1.87 \times 10^{-3}$~ms$^{-1}$. The time resolution in Hyper-Kamiokande is $\mathcal{O}$(ns)~\cite{Hyper-Kamiokande:2018ofw}, which is several orders of magnitude below the requirements for triangulation.

\subsection{DUNE}

DUNE will be built at Sanford, South Dakota in the US (see Table~\ref{tab:coordinates} for its latitude and longitude). DUNE will be a  liquid argon time projection chamber where scintillation light as well as drifted ionization signals will be used to precisely reconstruct timing and  topology of the recorded events. Its fiducial mass is expected to be $40$~kt. The main detection channel is the charged current  interaction between electron neutrinos and argon: $\nu_e + \Ar \rightarrow e^- + \K^\star$, whose cross section, $\sigma_{\mathrm{Ar}}$, is provided in Refs.~\cite{DUNE:2020zfm,GilBotella:2003sz}. The sub-leading neutral current  channel, $\nu + \Ar \rightarrow \nu + \Ar^\star$, is not included due to the uncertainties in its cross section. Other sub-leading channels that are not considered in this work  are $\bar{\nu}_e$ charged current interactions and elastic neutrino-electron scattering.

The main background in DUNE is expected to be coming from fast neutrons at a rate of $4$~s$^{-1}$ for an electron energy threshold of $6{\;\rm MeV}$, assuming no shielding and filtering~\cite{Li:2020ujl}. In addition, solar neutrinos are expected to contribute with a rate of $110$ per day with a $5{\;\rm MeV}$ threshold~\cite{Capozzi:2018dat}, and the spallation background for the same threshold is $4 \times 10^{-4}$~s$^{-1}$~\cite{Zhu:2018rwc}. Hence, in total, the background rate is assumed to be  $4.0017 \times 10^{-3}$~ms$^{-1}$. 

The main charged current channel proceeds through a number of different excited states, where the most important has a threshold neutrino energy of $5.888{\;\rm MeV}$~\cite{Li:2020ujl}. Hence, 
in order to control the background, we adopt ${E_{\rm thr}} = 12$~MeV  as neutrino energy threshold. While  $\epsilon_{\mathrm{DUNE}} = 86\%$ is used for the detection efficiency~\cite{DUNE:2015lol}. 

The resultant event rate for our BH forming collapse, $R_{\mathrm{DUNE}}$, is defined by substituting in \eref{eq:HKrate} the number of argon targets as well as the  efficiency of DUNE and the $\nu_e + \Ar$ 
cross section 
and it is  shown in the middle panel of \fref{fig:rate} as a function of time.  The time resolution for the time projection chamber is $0.6 {\;\rm ms}$, but this can be improved to sub ${\rm \mu s}$ resolution by using the scintillation light detected by the photon detector system~\cite{DUNE:2020zfm, DUNE:2018jwf}.

\subsection{IceCube Neutrino Observatory}

The IceCube Neutrino Observatory  consists of a cubic kilometer of instrumented ice at the South Pole; Table~\ref{tab:coordinates} reports its angular coordinates. This neutrino telescope is primarily intended for the detection of TeV--PeV  neutrinos and, hence, is  more sparsely instrumented than Hyper-Kamiokande and DUNE. Neutrinos in IceCube are primarily detected via IBD through the Cherenkov radiation emitted as the produced positron transverses the ice. Due to its sparse instrumentation, IceCube cannot detect individual neutrino events at MeV energies,  instead the neutrino burst from a BH forming collapse could be detected as a general increase in the background noise rate~\cite{IceCube:2011cwc}. 

Since there is no reconstructed energy, the IBD kinematic threshold  of $1.804{\; \rm MeV}$ is used as the threshold for detection. 
The effective mass of IceCube is $\sim 3.5{\;\rm Mton}$~\cite{IceCube:2011cwc}, and the time resolution  is a few ns~\cite{HeeremanvonZuydtwyck:2015mbs}. The IceCube event rate, $R_{\mathrm{IC}}$, is implemented by following the procedure outlined in Sec.~V of Ref.~\cite{Tamborra:2014hga}. While, by taking into account the detector dead time, the overall background rate is expected to be $1.48 \times 10^3$~ms$^{-1}$~\cite{IceCube:2011cwc}.
The resultant event rate for our BH forming collapse is shown in the bottom panel of Fig.~\ref{fig:rate} as a function of time.

%%%%%%%%%%%%%%%%%%%%%%%%%%%%%%%%%%%%%%%%%%%%%%%%%%
%%%%%%%%%%%%%%%%%%%%%%%%%%%%%%%%%%%%%%%%%%%%%%%%%%
\section{Timing the neutrino signal}
\label{sec:timing}
%%%%%%%%%%%%%%%%%%%%%%%%%%%%%%%%%%%%%%%%%%%%%%%%%%
%%%%%%%%%%%%%%%%%%%%%%%%%%%%%%%%%%%%%%%%%%%%%%%%%%

The angular location of the collapse of a massive star could be determined through triangulation, if the  neutrino signal happens to be recorded by several neutrino detectors in different geographical locations~\cite{Beacom:1998fj,Muhlbeier:2013gwa,Fischer:2015oma,Brdar:2018zds,Hansen:2019giq,Linzer:2019swe}. In fact, the time delay in the signal detected in  different neutrino telescopes allows to determine the neutrino arrival direction. 
We wish to compare the neutrino triangulation outcome by employing the  rise time~\footnote{The rise time of the neutrino rate coincides with the bounce time for Hyper-Kamiokande and IceCube; in the case of DUNE, this may not be the case given the pre-bounce $\nu_e$'s detectable in the ``no-mix'' scenario, see \fref{fig:rate}.}  (i.e., the approach most commonly adopted in the literature) and the cutoff of the neutrino curve at the moment of BH formation. To this end, in this section 
we evaluate  the accuracy and precision of timing  by taking into account the neutrino event rate expected in the neutrino telescopes introduced in Sec.~\ref{sec:detectors}. 
For the sake of simplicity, in this section, we assume that all detectors are at the same location, i.e.~there is no delay between the detectors due to the neutrino time of flight. 
  This calculation will be useful when correcting for the differences between the detectors (see Sec.~\ref{sec:triangulation}).

The expected event rate is the sum of the background rate and the event rate due to  neutrinos from the BH forming collapse, as shown in \fref{fig:rate}. 
In order to account for random coincidences, the pure background rate is added before and after the  neutrino signal from the  BH forming collapse.
The combined event rate is binned in $0.01{\; \rm ms}$ bins, and $10^4$ realizations of the neutrino event rate are drawn  for each detector by assuming Poisson statistics. For each realization, the rise time  and the end time of the neutrino signal are computed
 together with the time differences. From the set of realizations, the mean and uncertainty on each number are determined as detailed in this section.

\subsection{Determination of the rise time of the neutrino curve}\label{sec:timing_rise}

The  rise  times of the neutrino event rates in Hyper-Kamiokande and DUNE are determined by relying on  the first non-isolated detected event; i.e., the first event which is followed by another one in the time span $dt$. This method is similar to the one used in Ref.~\cite{Linzer:2019swe}, where $dt=15{\;\rm ms}$ was adopted. For our BH forming collapse  at $10{\; \rm kpc}$,  we assume  $dt = 1{\;\rm ms}$ for Hyper-Kamiokande and $dt = 0.66{\;\rm ms}$ for DUNE; this choice  minimizes the number of random coincidences and the effect of pre-bounce neutrinos. Even with our choice of $dt$,  the uncertainty in the determination of the rise time of the neutrino curve tends to be dominated by outliers, hence the first non-isolated events before $t_{\rm reject} = -10$~ms for Hyper-Kamiokande and $t_{\rm reject} = - 20$~ms for DUNE are rejected.

\begin{figure*}[tbp]
  \centering
    \includegraphics[width=0.5\textwidth]{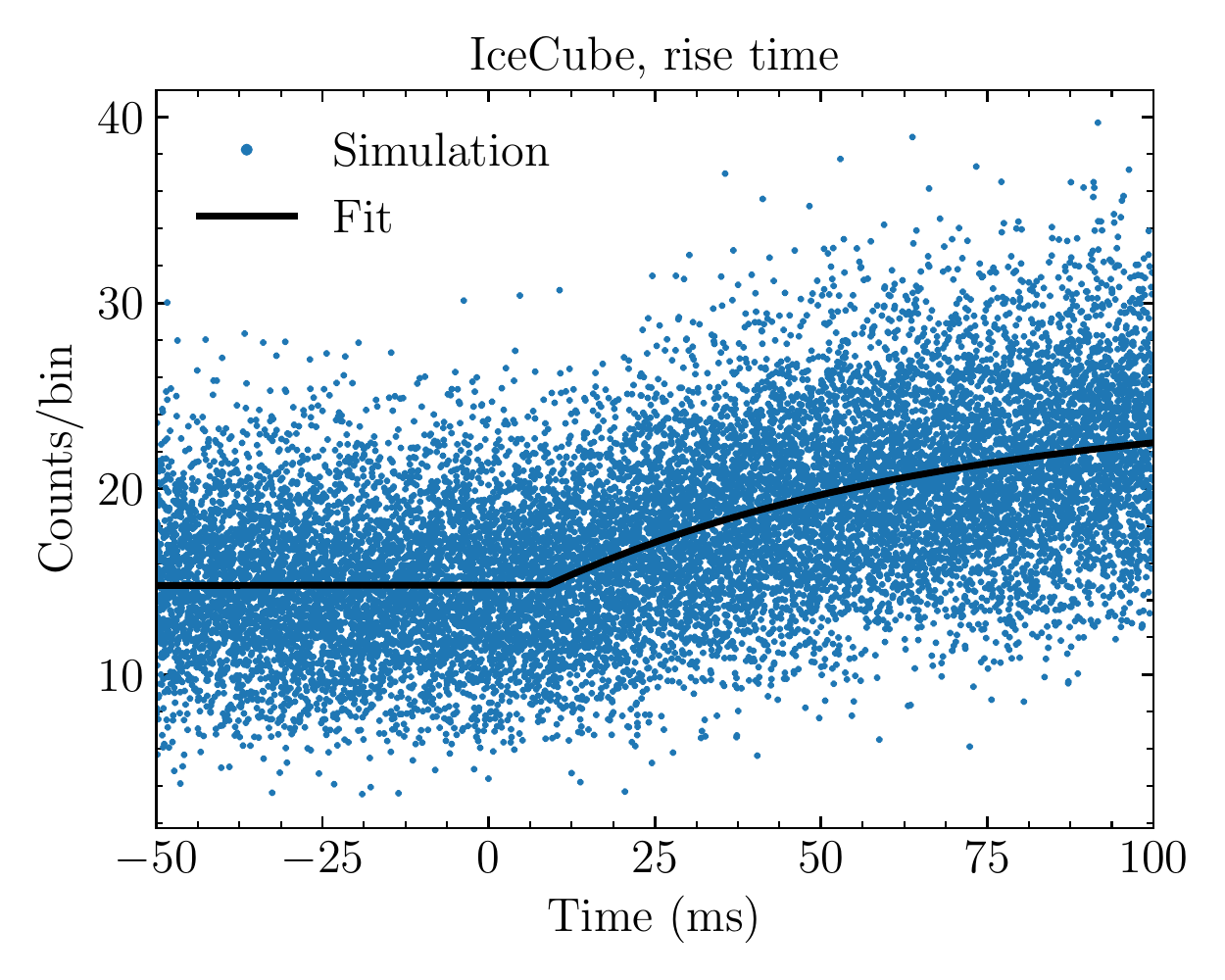}%
  \includegraphics[width=0.5\textwidth]{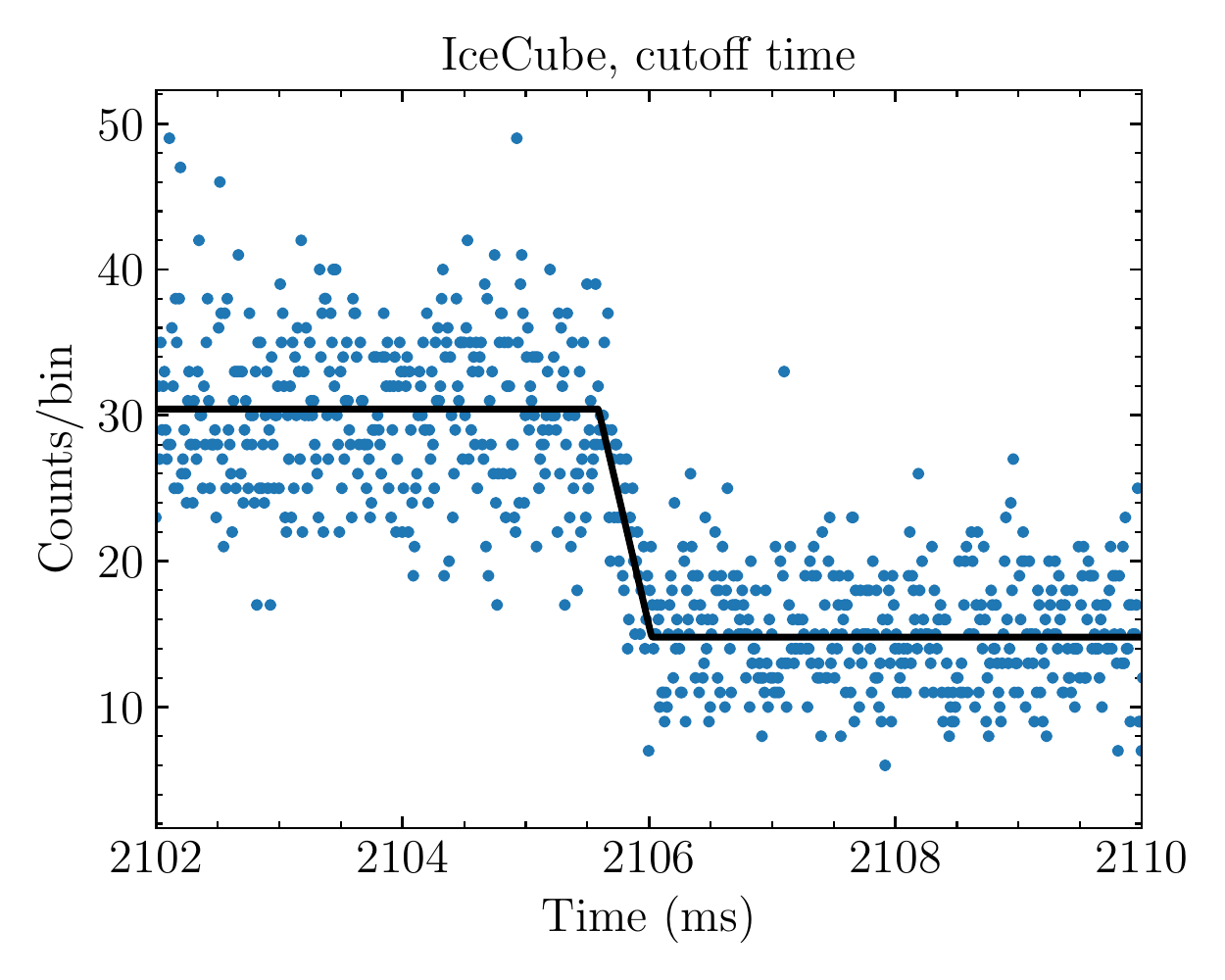}
  \caption{Illustrative example of the fitting procedure adopted for the IceCube event rate for the ``no-mix'' scenario for the rise time (left panel) and the cutoff time (right panels), see main text for details. The fits (solid lines) are in good agreement with the simulated expected signal (dots, obtained for one realization  where the number of events in each time bin is drawn from a Poisson distribution with mean equal to the event rate shown in the bottom panel of Fig.~\ref{fig:rate}).  This procedure allows to achieve an optimal determination of the time delays. Notice that the points in the left panel are offset on the y-axis by random numbers between -0.5 and 0.5 for readability.}
  \label{fig:ICendfit}
\end{figure*}

The approach described above  is not applicable to IceCube due to its large background rate. Instead, the following fit for the event rate is adopted~\cite{Halzen:2009sm}:
\begin{equation}\label{expfit2}
  R_{\rm exp, IC} =
  \begin{cases}
    0 &\text{if} \;t<t_0\\
    \tilde{R}_{\mathrm{IC}} \left(1 - e^{ -{(t-t_0)}/{\tau}}  \right)  &\text{otherwise}
 \end{cases}
\end{equation}
where $t_0$ is  the bounce time that we aim to determine and $\tau$ is a parameter fitted to the signal for each realization. This fitting function is applied between $-50{\;\rm ms}$  (i.e.~before the bounce, when the background dominates)  and the first $100{\;\rm ms}$ post bounce.  Then, the parameters $\tilde{R}_{\mathrm{IC}}$, $\tau$ and $t_0$ are fitted to the simulated signal for each event rate realization, as illustrated in the  left panel of Fig.~\ref{fig:ICendfit}. 

The mean and standard deviation of the rise time obtained over $10^4$ realizations are listed in the top of \tref{tab:arrival_bias} for Hyper-Kamiokande, DUNE and IceCube, for both mixing scenarios considered in this paper.  One can see that the rise time strongly depends on the mixing scenario and  the error on its determination  is larger in the ``no-mix'' scenario, because of  the slower rise time of the neutrino curve, as shown in \fref{fig:rate}. The  estimated rise time is lower in the ``full-mix'' scenario for Hyper-Kamiokande and IceCube, since the $\bar{\nu}_x$ signal starts slightly earlier than $\bar{\nu}_e$ and rises faster. The relatively large uncertainty in DUNE is mainly due to the pre-bounce neutrinos that produce a tail of estimated bounce times down to $t_{\rm reject}$. The peak due to the neutronization burst is not observable in  DUNE in the  ``full-mix'' scenario and, hence, the estimated mean rise time is larger. Since the pre-bounce neutrinos are not visible in the ``full-mix'' scenario in DUNE, the uncertainty is improved. This behavior could be avoided by tailoring the  timing method  to the mixing scenario---by using a smaller value of $dt$ in the ``no-mix'' scenario; however, this approach is not optimal given the current uncertainties on  flavor mixing.

The time difference for pairs of detectors is calculated by estimating the rise time from one realization in each detector and then by computing the difference between the bounce times measured in each of the two detectors. The  figures reported in \tref{tab:arrival_bias} are  the mean and standard deviation over $10^4$ realizations. One can see that the best determination of the time difference can be obtained by combining IceCube and Hyper-Kamiokande because of their very large statistics.

\begingroup
\setlength{\tabcolsep}{5pt}
\renewcommand{\arraystretch}{1.}
\begin{table}[tbp]
 
\caption{Estimation of the rise time  (top) and cutoff time (bottom) of the neutrino curve and related uncertainties for  Hyper-Kamiokande,  DUNE, and IceCube together with the time difference between pairs of neutrino telescopes in the two flavor mixing scenarios (absence of flavor mixing and full flavor mixing). Each of these figures has been obtained by computing the mean and standard deviation over $10^4$ realizations.  The time differences are computed by neglecting the neutrino time of flight delay (i.e.,~the detectors are assumed to be at the same location); hence they  correspond to the bias as defined in \eref{eq:bias}. }
  \begin{tabular}{l c c c}
    \hline\hline
    Method & Detectors & No-mix (ms) & Full-mix (ms) \\
    \hline
    \multirow{6}{*}{Rise time}
               & HK & $7.20 \pm 0.99$ & $3.30 \pm 0.41 $\\
               & DUNE & $-2.01 \pm 4.50$ & $11.59 \pm 3.50$\\
               & IC &  $9.15 \pm 1.13$ & $4.09 \pm 0.79$\\\\
               & IC-HK &$-1.95 \pm 1.51$ & $-0.79 \pm 0.88$ \\
               & HK-DUNE & $-9.21 \pm 4.62$ & $8.29 \pm 3.53$\\
               & DUNE-IC & $11.16 \pm 4.63$ & $-7.50 \pm 3.59$\\
 
    \hline
    \multirow{6}{*}{Cutoff time}
& HK & $2106.03 \pm 0.04$ & $2106.00 \pm  0.05$ \\
& DUNE & $2105.74 \pm 0.22$ & $2105.66 \pm 0.29$\\
& IC & $2106.10 \pm 0.04$ & $2106.10 \pm 0.07$\\\\
& IC-HK &$-0.07 \pm 0.06$ & $-0.10\pm 0.09$\\
& HK-DUNE & $-0.29 \pm 0.22$ & $-0.34 \pm 0.29$\\
& DUNE-IC & $0.36 \pm 0.22$ & $0.44 \pm 0.29$\\
    \hline\hline
  \end{tabular}
  \label{tab:arrival_bias}
\end{table}
\endgroup

\subsection{Determination of the cutoff time of the neutrino curve}\label{sec:timing_cutoff}

For Hyper-Kamiokande and DUNE, the determination of the cutoff time of the neutrino curve follows a procedure similar to the one adopted for the  rise time. The last non-isolated neutrino event is determined by selecting the last event, which is less than $dt$ after the second to last event. For Hyper-Kamiokande, $dt = 0.5{\;\rm ms}$ is used since a smaller value might discard events in the tail of the signal. For DUNE, $dt = 1.5{\;\rm ms}$ is adopted to maintain a low number of discarded events that are part of the signal as well as a low number of  outliers. 
If the last non-isolated event occurs at times larger than $t_{\rm reject} = 10\;{\rm ms}$ after the end of the signal, it is counted as an outlier and rejected.

For IceCube, the end of the event rate is  fitted through two plateaus connected by a linear function, as shown in the right panel of Fig.~\ref{fig:ICendfit}. 
The first plateau, the tail duration (expected to be around $0.5{\;\rm ms}$), as well as the end of the tail are determined by the fit.
The second plateau is taken to be the background rate determined after the tail. 
Note that if the actual position of the cutoff is used as the initial guess for the fit, the results are artificially pushed to this particular value. To avoid this, we draw the initial value from a Gaussian of width $0.05\;{\rm ms}$ centered on the actual cutoff.

The average cutoff times computed through this method are listed in the bottom part of \tref{tab:arrival_bias} for each of the considered neutrino telescopes as well as for their pairs. Note that the error in the determination of the cutoff time of the signal is worse for the ``full-mix'' scenario with respect to the ``no-mix'' case, as expected, given the smaller event rate (see Fig.~\ref{fig:rate}). Notably, since the number of expected events is larger at the end of the neutrino curve (due to the sustained accretion phase)  than for its initial rise, the determination of the end time is less sensitive to the flavor mixing scenario and it is on average more accurate.

%%%%%%%%%%%%%%%%%%%%%%%%%%%%%%%%%%%%%%%%%%%%%%%%%%
%%%%%%%%%%%%%%%%%%%%%%%%%%%%%%%%%%%%%%%%%%%%%%%%%%
\section{Triangulation}
\label{sec:triangulation}
%%%%%%%%%%%%%%%%%%%%%%%%%%%%%%%%%%%%%%%%%%%%%%%%%%
%%%%%%%%%%%%%%%%%%%%%%%%%%%%%%%%%%%%%%%%%%%%%%%%%%

  The angular location of a BH forming collapse  with respect  to two neutrino telescopes can be constrained  by measuring the time delay of the  signal measured in the two neutrino detectors. Detectors with a larger separation are expected to be less affected by the uncertainty in the determination of the delay times. Hence, the best chances for locating a BH forming collapse with high accuracy and precision are  expected  for detectors homogeneously distributed across the Earth. 
  In this section, we present our findings for the determination of the BH forming collapse angular location and its uncertainty by employing the rise time as well as the cutoff time of the neutrino curve and by taking into account the geographical displacement between the neutrino detectors as from \tref{tab:coordinates}.
  
  \subsection{Pinpointing the  angular location}
  The geographical location of the detectors adopted in this paper is reported  in \tref{tab:coordinates}; hence, the distance between each pair of detectors is:
   \begin{subequations}  \label{eq:separations}
  \begin{eqnarray}
      D_{\rm IC-HK} &=& 11374 {\;\rm km}\ ,\\
      D_{\rm HK-DUNE} &=& 8369 {\;\rm km}\ ,\\
      D_{\rm DUNE-IC} &=& 11746 {\;\rm km}\ .
  \end{eqnarray}
    \end{subequations}
 These large distances demonstrate that all three detectors are well separated, and the maximal time of flight for a relativistic neutrino traveling between two detectors is about $\mathcal{O}(30){\;\rm ms}$ for all three detector combinations. 

  Locations on the sky can be defined in the equatorial coordinate system~\footnote{In the equatorial coordinate system, right ascension $\alpha$ corresponds to longitude and declination $\delta$ corresponds to latitude however, it does not rotate with the Earth.
  At vernal equinox, the sun is at $\alpha=0$, $\delta=0$. See also~\cite{Muhlbeier:2013gwa}.}, which is fixed against background stars. We assume that the BH forming collapse is observed at noon UTC on vernal equinox. For a BH forming collapse at right ascension $\alpha$ and declination $\delta$, the travel direction of the  neutrinos can be calculated as
    \begin{equation}
      \label{eq:nvector}
      \mathbf{n} = - (\cos\alpha \cos\delta, \sin\alpha \cos\delta, \sin\delta)^T .
    \end{equation}
    The time delay between the detectors $i$ and $j$ is
    \begin{equation}
      \label{eq:timedelay}
      t_{i,j} = \frac{\mathbf{r}_j - \mathbf{r}_i}{c} \cdot \mathbf{n}\ ,
    \end{equation}
    where $\mathbf{r}_i$ is the position of the detector $i$ in Cartesian 
    coordinates, and $c$ is the speed of light.

The measure of the neutrino  time delays is  not only determined by the location of the BH forming collapse with respect to the neutrino detectors, but also by the method employed for the timing. In fact, the different detector materials, sizes, and technologies lead to time estimates that deviate beyond statistical fluctuations---this uncertainty is enclosed in a bias term, and the measured time delay is defined as~\cite{Linzer:2019swe}:
  \begin{equation}
    \label{eq:bias}
    t_{i,j}^{\rm measured} = t_{i,j}^{\rm true} + B_{i,j}\ .
  \end{equation}
The bias $B_{i,j}$ between the detectors $i$ and $j$ can be estimated as the mean time differences in \tref{tab:arrival_bias}, where $t_{i,j}^{\rm true} = 0$. 
The measured time delays defined as in Eq.~(\ref{eq:bias})  and assuming a BH forming collapse at the Galactic Center detected at vernal equinox at noon are listed in  \tref{tab:arrival}.
  The trend is very similar to what was found in \tref{tab:arrival_bias}, and the physical explanations are the same.
By subtracting the biases, the expected time delays are recovered with good precision (results not shown here).

\begingroup
\setlength{\tabcolsep}{5pt}
\renewcommand{\arraystretch}{1.}
\begin{table}
\caption{Same as \tref{tab:arrival_bias}, but  obtained by taking into account the different geographical locations of the neutrino telescopes and a BH forming collapse occurring at the Galactic Center.}
\begin{tabular}{l c c c}
\hline\hline
Method & Detectors & No-mix (ms) & Full-mix (ms) \\
\hline
  \multirow{6}{*}{Rise time}
& HK & $32.86 \pm 1.01$ & $28.96 \pm 0.41 $ \\
& DUNE & $2.34 \pm 4.46$ & $16.03 \pm 3.52$ \\
& IC &$9.15 \pm 1.13$ & $4.09 \pm 0.79$ \\\\
& IC-HK &$23.71 \pm 1.50$ & $24.87 \pm 0.89$ \\
& HK-DUNE & $-30.52 \pm 4.56$ & $-12.93 \pm 3.54$ \\
& DUNE-IC & $6.81 \pm 4.61$ & $-11.94 \pm 3.60$ \\
\hline

  \multirow{6}{*}{Cutoff time}
& HK & $2131.69 \pm 0.04$ & $2131.66 \pm 0.05 $ \\
& DUNE & $2110.10 \pm 0.21$ & $2110.02 \pm 0.28$ \\
& IC & $2106.10 \pm 0.04$ & $2106.10 \pm 0.07$ \\\\
& IC-HK &$25.58 \pm 0.06$ & $25.56 \pm 0.09$ \\
& HK-DUNE & $-21.59 \pm 0.22$ & $-21.64 \pm 0.29$ \\
& DUNE-IC & $-4.00 \pm 0.22$ & $-3.92 \pm 0.29$ \\
\hline\hline
\end{tabular}
\label{tab:arrival}
\end{table}
\endgroup

The time delay between a pair of detectors allows to determine the angle $\theta$ spanned by the two lines connecting the detectors and the location of the BH forming collapse in  the sky. This angle defines a cone in the sky. Using  three detectors, three cones would intersect in two points, while  a unique location can be determined with four detectors. In this paper, we focus on combining three detectors together in order to show the potential of the method employing the cutoff time of the neutrino curve.

\subsection{Determination of the angular uncertainty}

  The uncertainty on the angle $\theta$ for each pair of detectors can be determined without 
  full triangulation~\cite{Beacom:1998fj, Hansen:2019giq}. In fact, for a given pair of detectors, 
  \begin{equation}
    \label{eq:costheta}
    \cos\theta_{i,j} = \frac{t_{i,j}^{\rm true}}{D_{i, j}}\ ,
  \end{equation}
  where  $D_{i, j}$ is the distance between the two detectors (see  Eqs.~\ref{eq:separations}). 
The resulting uncertainty on the angle is~\cite{Beacom:1998fj, Hansen:2019giq}:
  \begin{equation}
    \label{eq:deltatheta}
    \delta(\theta_{i,j}) \approx
    \begin{cases}
     \delta(\cos\theta_{i,j})/\sin\theta_{i,j} & {\rm if \;} \sin\theta_{i,j} > \sqrt{\delta(\cos\theta_{i,j})}\ ,\\
      \sqrt{2\delta(\cos\theta_{i,j})} & {\rm for\;} \theta_{i,j} \ll \delta(\cos\theta_{i,j})\ ,
    \end{cases}
  \end{equation}
  which allows to determine the maximal and minimal uncertainties on the angular location of the BH forming collapse.

 In order to derive the uncertainty on $t_{i,j}^{\rm true}$,     the bias uncertainty and  the one related to $t_{i,j}^{\rm measured} $  are added in quadrature (see Eq.~(\ref{eq:bias})):
  \begin{equation}
    \label{eq:totalsigma}
    \sigma_{\rm true}^2 = \sigma_{\rm bias}^2 + \sigma_{\mathrm{measured}}^2\ .
  \end{equation}
This approach   leads to an   increase of the uncertainty on $t_{i,j}^{\rm true} $ with respect to  $t_{i,j}^{\rm measured} $  by approximately $\sqrt{2}$. 
Although the employment of the  true bias might provide an optimistic pointing accuracy,  our main focus is on pointing precision and, for that, \eref{eq:totalsigma} is quite conservative~\cite{Linzer:2019swe}.   

The minimal and maximal uncertainties, $\delta(\theta_{i,j})$,  computed through this approach are reported in \tref{tab:angle} for the method employing the rise time of the neutrino curve and  the one relying on the cutoff time. By using the cutoff time rather than the rise time, one can see that the minimal uncertainty is significantly below $1$~deg.

\begingroup
\setlength{\tabcolsep}{5pt}
\renewcommand{\arraystretch}{1.}
\begin{table}[tbp]
\caption{Maximum and minimum angular uncertainties ($\delta(\theta_{i,j})$, see Eq.~\ref{eq:deltatheta}) derived by applying the rise  and cutoff  time methods for the  pairs of detectors in our two mixing scenarios. The cutoff time method allows to achieve an order of magnitude improvement on the angular uncertainty. }
\begin{tabular}{l c c c}
\hline\hline
Method & Detectors & No-mix (deg) & Full-mix (deg) \\
\hline
\multirow{4}{*}{Rise time} 
& IC-HK    & $[3.22, 19.20]$ & $[1.90, 14.74]$ \\
& HK-DUNE  & $[13.32, 39.07]$ & $[10.26, 34.29]$ \\
& DUNE-IC  & $[9.55, 33.08]$ & $[7.44, 29.20]$ \\

\hline
\multirow{4}{*}{Cutoff time}  
& IC-HK  &$[0.12, 3.76]$ & $[0.18, 4.59]$ \\
& HK-DUNE  & $[0.64, 8.56]$ & $[0.84, 9.80]$ \\
& DUNE-IC  & $[0.46, 7.24]$ & $[0.60, 8.29]$ \\

\hline\hline
\end{tabular}
\label{tab:angle}
\end{table}
\endgroup

\subsection{Combining  multiple detectors}

When combining the time differences between $n$ different detectors, correlations between the  detector combinations must be accounted for. This is done by including the covariance matrix $V$. In analogy with \eref{eq:totalsigma}, we account for the bias correction by using
\begin{equation}
  \label{eq:totalV}
  V_{\rm true} = V_{\rm bias} + V_{\rm measured}\ ,
\end{equation}
where each matrix is of dimension $m \times m$, with $m$  given by the binomial coefficient $m=\binom{n}{2}$. The entries in the covariance matrices correspond to the time delay vector 
\begin{equation}
  \label{eq:tdelay}
  \mathbf{t} = (t_{1,2}, t_{1,3}, \dots, t_{1,n}, t_{2,3}, \dots, t_{n-1,n})^T \ .
\end{equation}
In order to determine the preferred regions on the sky, we rely on the chi-square:
\begin{equation}
  \label{eq:chisquared}
  \chi^2(\alpha,\delta) = (\mathbf{t}(\alpha,\delta) - \mathbf{t}^{\rm true} )^T\ V_{\rm true}^{-1}\ (\mathbf{t}(\alpha,\delta) - \mathbf{t}^{\rm true} )\ ,
\end{equation}
where $\mathbf{t}^{\rm true}$ 
is computed using \eref{eq:bias} with the values from Tables~\ref{tab:arrival} and \ref{tab:arrival_bias} 
and $\mathbf{t}(\alpha,\delta)$ is the time delay vector computed by assuming that the BH forming collapse has right ascension $\alpha$ and declination $\delta$.

\begin{figure*}[tbp]
  \centering
    \includegraphics[width=0.5\textwidth]{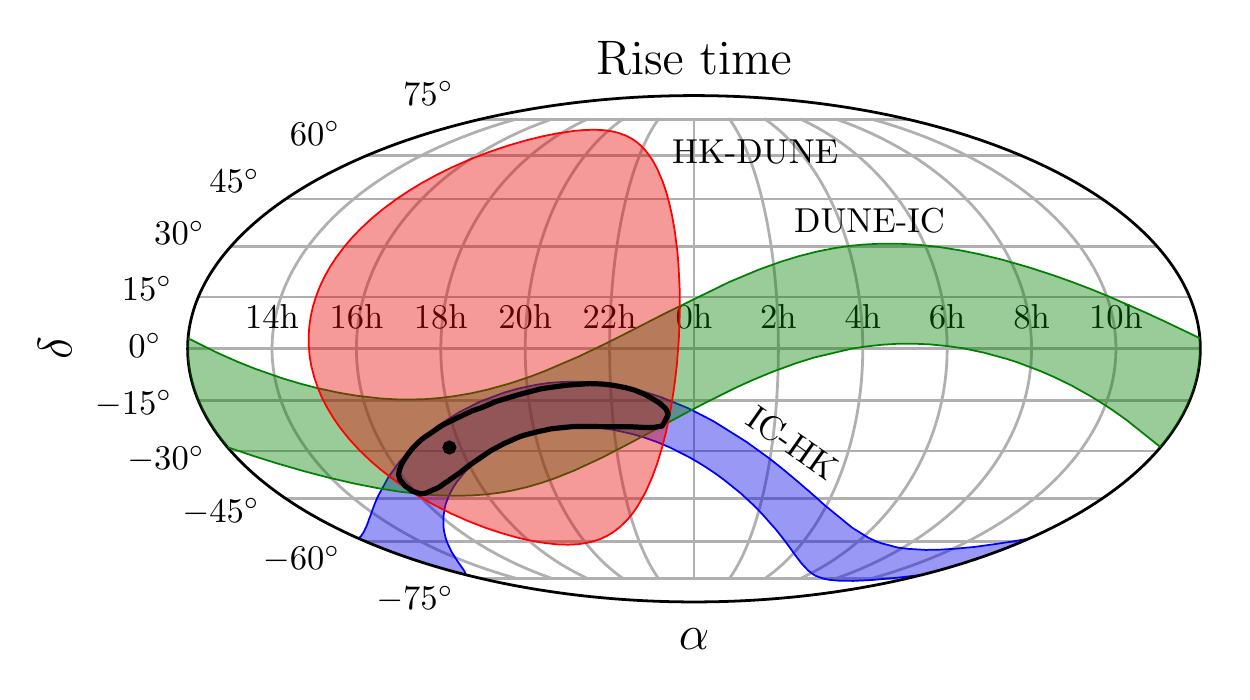}%
    \includegraphics[width=0.5\textwidth]{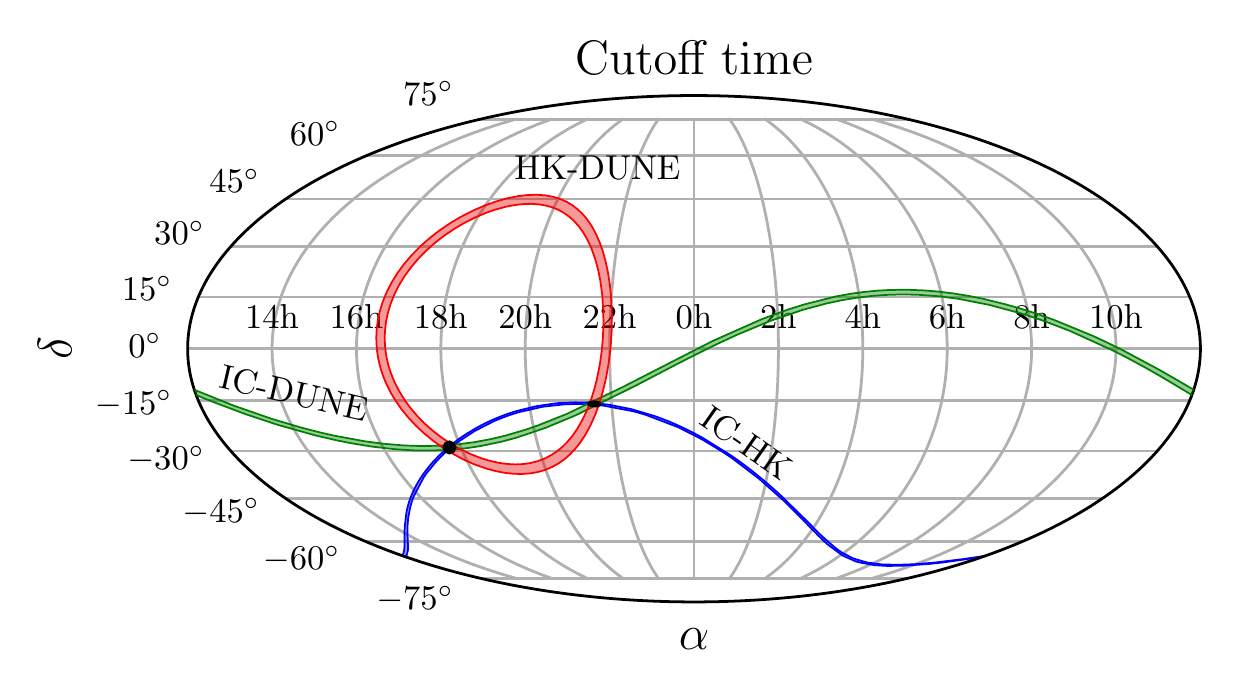}
    \caption{Uncertainty bands  ($1\sigma$)  of the location of our BH forming collapse at $10$~kpc on a skymap  in the ``no-mix'' scenario  obtained by employing the method based on the rise time of the neutrino curve (left panel) and the cutoff time of the neutrino curve (right panel). The true location of the BH forming collapse is marked with a black dot. The pointing precision dramatically improves for the cutoff time method. 
    }
  \label{fig:maps}
\end{figure*}

For two detectors, there is only one entry in $\mathbf{t}$, and the matrix product reduces to a simple scalar product. Assuming no correlations, \eref{eq:chisquared} reduces to a sum over such scalar product~\cite{Muhlbeier:2013gwa, Brdar:2018zds, Linzer:2019swe}.
However, when more detectors are included, the time delays are correlated by definition, and using a diagonal covariance matrix  underestimates the uncertainty in determining the direction.

The sky regions constrained by  combining two and three detectors using the rise times and cutoff times  are shown in the left and right panels of \fref{fig:maps}, respectively. The colored bands represent the regions of the sky constrained by  using the indicated two-detector combinations, whereas the black contours represent the combinations of three detectors. As expected, the pointing precision of the cutoff times is superior to that of the rise times.

%%%%%%%%%%%%%%%%%%%%%%%%%%%%%%%%%%%%%%%%%%%%%%%%%%
%%%%%%%%%%%%%%%%%%%%%%%%%%%%%%%%%%%%%%%%%%%%%%%%%%
\section{Dependence of the pointing precision on the distance of the black hole forming collapse from Earth}
\label{sec:distance}
%%%%%%%%%%%%%%%%%%%%%%%%%%%%%%%%%%%%%%%%%%%%%%%%%%
%%%%%%%%%%%%%%%%%%%%%%%%%%%%%%%%%%%%%%%%%%%%%%%%%%

The flux of neutrinos from a BH forming collapse scales with  the inverse distance squared, hence affecting the determination of the angular location of a BH forming collapse.  In this section, we investigate  the dependence of the timing and  pointing precision from   the distance at which the BH forming collapse is located with respect to  Earth.

\subsection{Dependence of the pointing precision on the distance:  rise-time method}

The results presented so far are based on the assumption of a  BH forming collapse  at $d=10\;{\rm kpc}$ from Earth. However, as shown in Eq.~\eqref{eq:flux}, the neutrino flux scales as $d^{-2}$. 
Consequently, we should expect that the typical delay between the first event and a following event increases with $d^2$.  
When determining the rise time in Hyper-Kamiokande and DUNE, the time span $dt$ and the rejection time $t_{\rm reject}$ introduced in Sec.~\ref{sec:timing_rise} should scale as  functions of $d$.  For Hyper-Kamiokande, the event rate is high enough that we do not need to increase $t_{\rm reject}$, and we keep it constant at $-10\;{\rm ms}$. Whereas, for DUNE, we use a rejection time  $t_{\rm reject} = -\left({d}/{10\;{\rm kpc}}\right)^2 \times 20 \; {\rm ms}$ for distances above $10\;{\rm kpc}$ and  $t_{\rm reject} = -20\;{\rm ms}$ for smaller distances.
The time span $dt$ is simply scaled with $d^2$ to match its value at $10\;{\rm kpc}$. 

The rise time in IceCube can be determined by adapting the fitting function introduced in \eref{expfit2} to the signal (see also \fref{fig:ICendfit}). This approach continues to give a good estimate of the timing up to  $60\;{\rm kpc}$.  Above this distance, the uncertainty becomes so large that no directional information can be extracted through triangulation. 

The minimal and maximal angular uncertainties as a function of the distance of the BH forming collapse derived from the rise time are shown in the left panel of \fref{fig:SNdist}. The uncertainty increases gradually until a distance of $25\;{\rm kpc}$ above which the timing in DUNE degrades rapidly. Notice that, although the bands for IC-DUNE and HK-DUNE appear to narrow in the proximity of $25\;{\rm kpc}$, the vertical width does not shrink significantly while the horizontal one scales as a function of the distance. 
On the other hand, the combination of IceCube and Hyper-Kamiokande displays  a  gradual increase in uncertainty until $60\;{\rm kpc}$ where the minimum uncertainty on the angle is $\sim 50^\circ$ corresponding to a very large fraction of the sky.

\subsection{Dependence of the pointing precision on the distance:  cutoff-time method}

The determination of the cutoff time for a BH forming collapse at $10\;{\rm kpc}$ has been introduced in Sec.~\ref{sec:timing_cutoff}. As $d$ varies, 
we use $dt=0.5\;{\rm ms}$ for Hyper-Kamiokande and $dt=1.5\;{\rm ms}$ for DUNE. As for $t_{\rm reject}$, we use  $t_{\rm reject} = 10$~ms  for Hyper-Kamiokande independently of $d$, since the event rate is so large that it allows for an excellent discrimination of the signal from the background even for large $d$. 
On the other hand, for DUNE, we use $t_{\rm reject} = \left({d}/{10\;{\rm kpc}}\right)^2 \times 10 \;{\rm ms}$ and fix  $t_{\rm reject} = 125\;{\rm ms}$ for $d$ providing an otherwise larger $t_{\rm reject}$.

\begin{figure*}[tbp]
  \centering
  \includegraphics[width=0.5\textwidth]{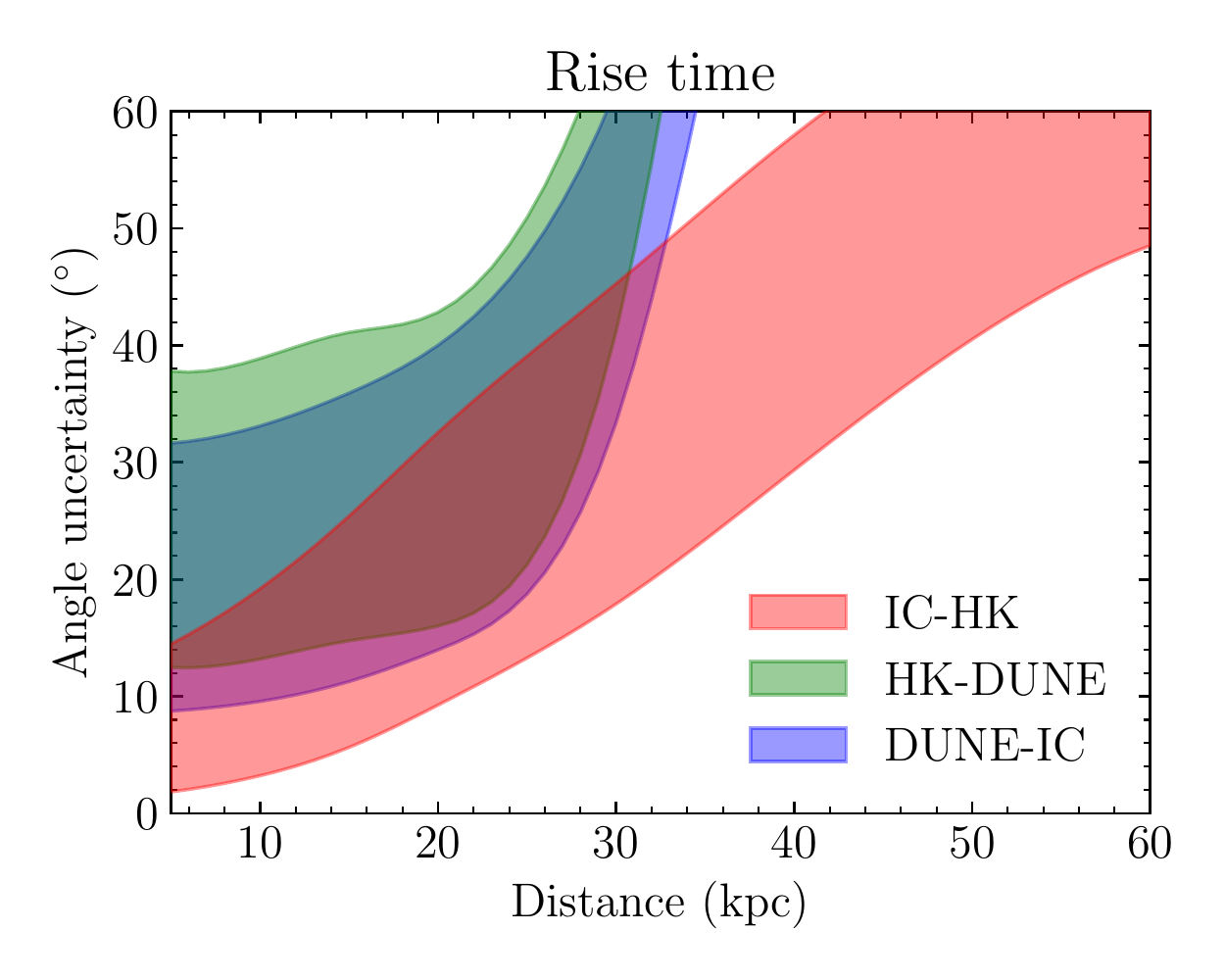}%
  \includegraphics[width=0.5\textwidth]{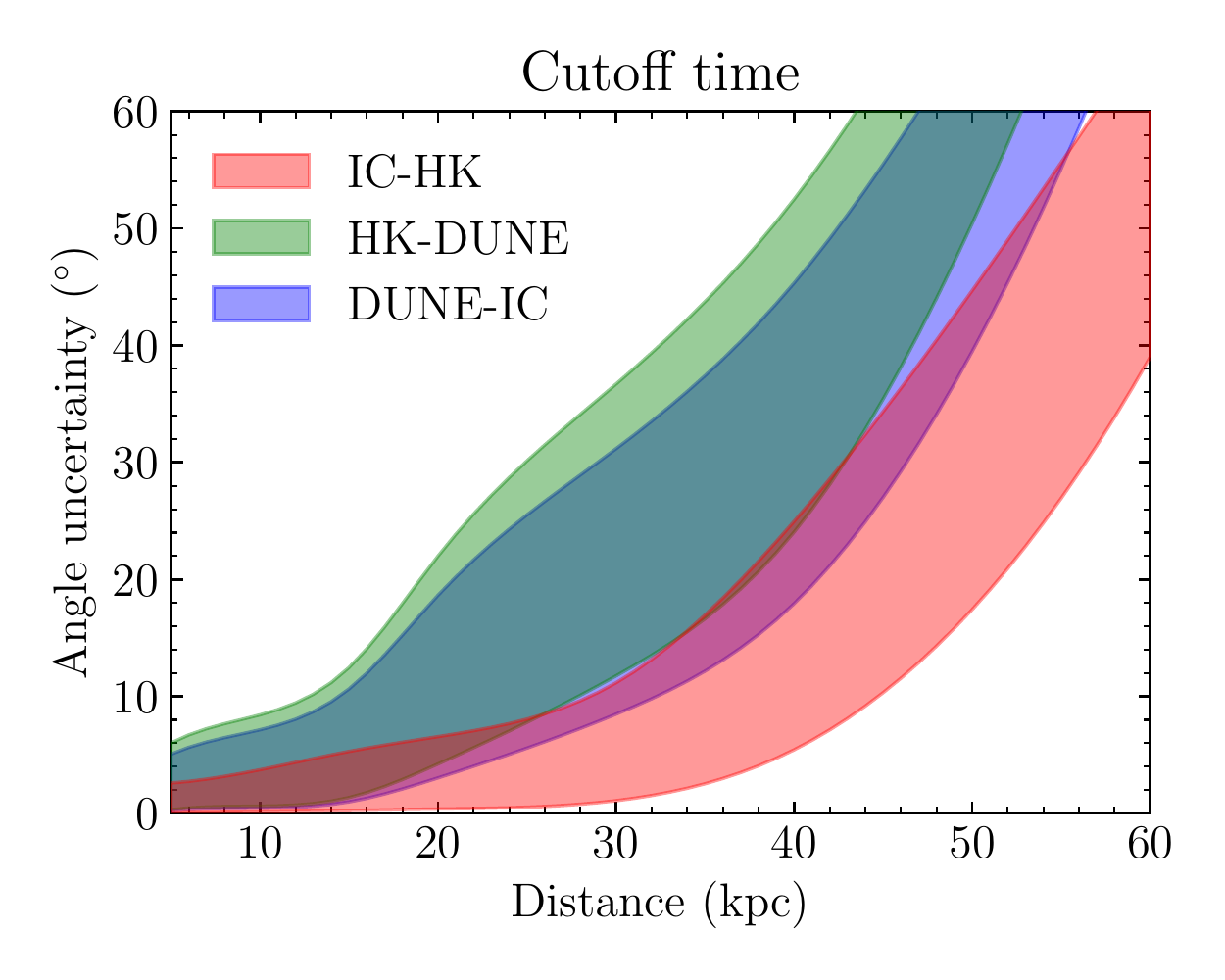} 
  \caption{Angular uncertainty range (the uncertainty band represents the maximal and minimal angular uncertainties as computed from \eref{eq:deltatheta})  on the determination of the BH forming collapse location in the ``no-mix'' scenario as a function of the distance for the rise time method (left panel) and the cutoff time method (right panel). The minimum and maximum angular uncertainties are calculated using \eref{eq:deltatheta}, and the uncertainty is corrected for the effect of the bias by multiplying with $\sqrt{2}$ (see Eq.~\ref{eq:totalsigma}). The pointing accuracy is  significantly better when using the cutoff time method (right). In addition, due to their large statistics, the combination of IceCube and Hyper-Kamiokande allows to achieve a better performance. }
  \label{fig:SNdist}
\end{figure*}

As for the  cutoff fit for IceCube (see \fref{fig:ICendfit}),
we cannot assume that we  a priori  know the time of the cutoff within $\pm 0.05\;{\rm ms}$ for any $d$, hence we draw the initial guess from a wider distribution. An appropriate choice for the width is
 $   \sigma_{\rm draw} = 25 {R_{\rm IC, bckgd}}/{R_{\rm IC}^2}$, 
where $R_{\rm IC, bckgd}$ and $R_{\rm IC}$ are the background and signal rates respectively. Due to the linear tail, we require $\sigma_{\rm draw}>0.05\;{\rm ms}$.
  At $d=10\;{\rm kpc}$, a period with pure background of $10\;{\rm ms}$ is added after the signal. 
  The period with background is increased to $3\;\sigma_{\rm draw}$ when $3\;\sigma_{\rm draw} > 10\;{\rm ms}$. 
  
The minimal and maximal angular uncertainties from the cutoff time method are shown in the right panel of \fref{fig:SNdist}. 
At distances up to $\sim 15\;{\rm kpc}$, the maximal uncertainty is below $10^\circ$. At larger distances the precision decreases notably and, as for the rise time method, the timing in DUNE  degrades fast. 
The combination of IceCube and Hyper-Kamiokande provides the highest precision at all distances. For BH forming collapses throughout the Milky Way (up to $23\;{\rm kpc}$) the precision is well below $10^\circ$. At larger distances, the noise in IceCube leads to larger uncertainties, and at $\sim 60\;{\rm kpc}$ the precision is comparable  to the one obtained with the  rise time method (see left panel of \fref{fig:SNdist}).

  The uncertainty achieved by combining  all three detectors cannot be directly  extrapolated from the bands for two detector combinations in \fref{fig:SNdist}, but the general trends are similar for two reasons. First, the width of the combined region will never be larger than the smallest uncertainty of the two detector combinations. Second, in the region where DUNE has no sensitivity (rise time beyond $\sim 30$ kpc) the only constraint will come from the combination of IceCube and Hyper-Kamiokande, and hence this gives a very accurate picture of the full three detector uncertainty. Consequently the width of the bands in \fref{fig:SNdist} are a good proxy for the uncertainty in the three detector case, although precise numbers require additional calculations that go beyond the scope of this work.

%%%%%%%%%%%%%%%%%%%%%%%%%%%%%%%%%%%%%%%%%%%%%%%%%%
%%%%%%%%%%%%%%%%%%%%%%%%%%%%%%%%%%%%%%%%%%%%%%%%%%
\section{Discussion}
\label{sec:discussion}
%%%%%%%%%%%%%%%%%%%%%%%%%%%%%%%%%%%%%%%%%%%%%%%%%%
%%%%%%%%%%%%%%%%%%%%%%%%%%%%%%%%%%%%%%%%%%%%%%%%%%

The employment of the cutoff time method is clearly promising for the determination of the angular location of a naked stellar-mass BH, in agreement with the findings of Refs.~\cite{Muhlbeier:2013gwa,Brdar:2018zds,Hansen:2019giq}. One of the reasons  is the high event rate due to the long-lasting accretion phase, preceding the instant of BH formation. This, combined with the steepness of  the tail of the neutrino curve, allows for a better discrimination of the signal with respect to the background than in the case of the rise time method. In the following, we discuss some of the unknowns potentially affecting our analysis.

\begin{figure}[tbp]
  \centering
  \includegraphics[width=0.45\textwidth]{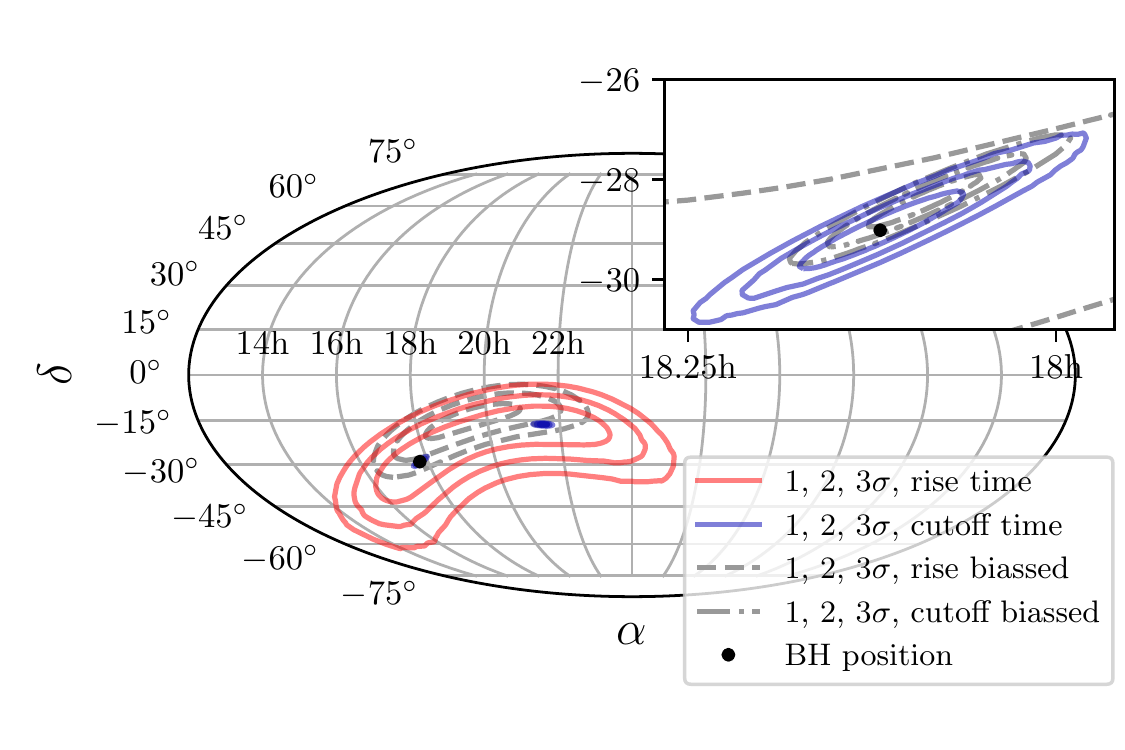}
  \caption{Impact of the bias on the determination of the location of the BH forming collapse in the ``no-mix'' scenario. The true location of the BH forming collapse is marked with a black dot. The $1, 2,$ and $3\ \sigma$ confidence level contours are plotted in gray for the cases without bias correction, in red for the rise time method, and in blue for the cutoff time method; the latter is magnified in the plot inset. Correcting for the the bias affects the pointing accuracy and it should not be neglected. The red and blue $1\sigma$ contours correspond to the black lines in \fref{fig:maps}.}
  \label{fig:maps_bias}
\end{figure}

\begin{itemize}
\item {\it Astrophysical uncertainties}. If the newly-formed BH is not naked, a neutrino echo could affect the tail of the neutrino curve because of the non-negligible supersonic accretion flow surrounding the central compact object~\cite{Gullin:2021hfv}. Hence,  one should expect a smearing of the tail of the neutrino curve and a less precise determination of the BH forming collapse location. Still, we expect that the cutoff time method will perform better than the rise time method due to the sharper transition between the time bins where the event rate is  dominated by the signal and the time bins dominated by the background.

The modeling of the tail of the neutrino curve has been inspired by Ref.~\cite{Baumgarte:1996iu} for our benchmark BH forming collapse. Of course, one should keep in mind eventual variations on the steepness of the tail due to current uncertainties in the theoretical modelling of BH forming collapses and the lack of multi-dimensional hydrodynamical simulations tracking the instant of BH formation.

In addition to the linear tail,  an exponential component with a decay time of $\mathcal{O}(0.05)\;{\rm ms}$~\cite{Baumgarte:1996iu,Beacom:2000qy,Wang:2021elf}, due to the neutrino trajectories being bent by the gravitational field, should be expected. This time scale is comparable to the precision that we obtain with Hyper-Kamiokande and IceCube, and should not increase the overall  uncertainty by more than a factor of two. Hence, we expect that the impact of this correction on our results should be minor.

\item {\it Neutrino mixing}. Because of the current uncertainties involved in  the modeling of the neutrino flavor evolution in neutrino-dense media~\cite{Mirizzi:2015eza,Duan:2010bg,Tamborra:2020cul}, we have chosen two extreme scenarios: absence of flavor mixing or full flavor conversion. Importantly, the rise time method strongly depends  on the mixing scenario because of the different rise of the electron and non-electron neutrino curves (Fig.~\ref{fig:rate}); whereas, the cutoff method is less sensitive to it, see Table~\ref{tab:angle}, due to the anyway large event rate. In this sense, despite the astrophysical uncertainties, the  cutoff time method is more promising than the rise time method.

\item {\it Proposed methodology}. 
Our results are in agreement with the findings of Refs.~\cite{Muhlbeier:2013gwa,Brdar:2018zds}. However,  despite comparable theoretical  input for the neutrino curve with respect to Ref.~\cite{Muhlbeier:2013gwa}, our statistical method allows to achieve an accuracy below ms even for the rise time method (see Tables~\ref{tab:arrival_bias} and \ref{tab:angle}), while Ref.~\cite{Muhlbeier:2013gwa}  assumed a few ms precision. 
Our work differs from Ref.~\cite{Brdar:2018zds} since we focus on next generation neutrino detectors and  optimize the  statistical  technique as well as the detection strategy.  

\item {\it Bias.}
  A word of caution should be added concerning the bias often neglected in the existing literature, see e.g.~Refs.~\cite{Muhlbeier:2013gwa, Brdar:2018zds}.
  The true bias of the time differences is estimated by the mean values in \tref{tab:arrival_bias}. For the rise time method, the bias strongly depends on the mixing scenario. As a consequence, it is difficult to provide reliable theoretical input for determining the bias. On the contrary, the bias for the  cutoff time method is almost independent of mixing since the signal in all flavors is cut off simultaneously. 
  
  A data driven bias could be  determined as well~\cite{Linzer:2019swe}. This does not rely on any model assumptions, but also cannot account for the full bias. 
  Since for the cutoff time method, the bias is largely independent on the mixing scenario,  the true bias determined in \tref{tab:arrival_bias} is more robust to our purpose than a data driven one.

  In order to highlight the impact of the bias on the determination of the angular location of a BH forming collapse, \fref{fig:maps_bias} shows the uncertainty contours for the rise and cutoff time methods  for the cases with bias correction as in \fref{fig:maps} (red and blue contours) and the ones without bias correction (gray contours), i.e.~when $t_{i,j}^{\rm measured}$ is used to estimate the BH location in \eref{eq:chisquared}. One can clearly see that, without including the bias, the best fit is shifted with respect to the true location of the stellar collapse and the uncertainty bands are underestimated. 
  
  \item {\it Experimental uncertainties.} Despite the fact that we have developed an improved strategy to determine the cutoff time of the neutrino curve, our method remains sensitive to our knowledge of the detector backgrounds. For example, the possibility of further reducing the backgrounds (e.g.~through gadolinium enrichment in Hyper-Kamiokande) as well as a better knowledge of the backgrounds (e.g.~in DUNE) could surely improve our findings concerning the determination of the angular accuracy on the location of the BH forming collapse. In addition, the employment of the neutral current channel in DUNE could boost the event rate, allowing a better timing performance, especially as the distance of the BH forming collapse increases.
  
  \end{itemize}
  
%%%%%%%%%%%%%%%%%%%%%%%%%%%%%%%%%%%%%%%%%%%%%%%%%%
%%%%%%%%%%%%%%%%%%%%%%%%%%%%%%%%%%%%%%%%%%%%%%%%%%
\section{Conclusions}
\label{sec:conclusions}
%%%%%%%%%%%%%%%%%%%%%%%%%%%%%%%%%%%%%%%%%%%%%%%%%%
%%%%%%%%%%%%%%%%%%%%%%%%%%%%%%%%%%%%%%%%%%%%%%%%%%

The collapse of a massive star could lead to BH formation. In this case, neutrinos  could carry precious input to locate the BH forming collapse in the sky. In this work, we explored the 
possibility of improving on the determination of the angular location of a BH forming collapse through triangulation pointing, under the assumption that accretion in the surrounding of the central compact object is halted and naked BH formation occurs. 

By forecasting the neutrino event rate from a BH forming collapse expected in the upcoming Hyper-Kamiokande and DUNE as well as the existing IceCube Neutrino Observatory, we show that focusing on the tail of the neutrino curve rather than on its bounce time, as often considered in the literature,  allows to achieve at least an order of magnitude  better determination of the angular location of the stellar collapse. This is due to the   larger event rate expected during the accretion phase preceeding BH formation, which is more easily distinguishable from the detector backgrounds. Notably, despite  the existing uncertainties in the modeling of neutrino mixing in the stellar core,  we find that the employment of the cutoff time of the neutrino signal for triangulation pointing is less sensitive to flavor mixing. 

This work also highlights the importance of the bias in the determination of the angular location of the stellar collapse. In fact, neglecting it could lead to a wrong estimation of the true location  of the BH forming collapse as well as to underestimate the uncertainty bands. Given the similar behavior for all flavors at cutoff, the bias correction is also less sensitive to the mixing scenario when the cutoff time method is employed.

Our  improved strategy for  triangulation pointing of BH forming collapses contributes to the growing set of  real-time strategies to improve  the   pointing capabilities through multi-messenger data. However, work remains to be done to pin down the theoretical uncertainties in the modeling of the neutrino signal at BH formation in multi-dimensions as well as to improve our understanding of the detector responses and backgrounds.

\acknowledgments
We thank  Thomas Janka, Kate Scholberg and Anna Suliga for insightful discussions. This project has received funding from the  Villum Foundation (Projects No.~13164 and No.~37358), the European Union's Horizon 2020 research and innovation program under the Marie Sklodowska-Curie grant agreement No.~847523 (``INTERACTIONS''), the Danmarks Frie Forskningsfonds (Project No.~8049-00038B), and the Deutsche Forschungsgemeinschaft through Sonderforschungbereich SFB~1258 ``Neutrinos and Dark Matter in Astro- and Particle Physics'' (NDM).

%\appendix

\bibliography{literature}

\end{document}